\documentclass[twocolumn,tighten]{aastex62}
\usepackage{float}
\usepackage{graphicx}
\usepackage{natbib}
\usepackage{txfonts}
\usepackage[flushleft]{threeparttable}
\usepackage{xcolor}

\graphicspath{{./}{figures/}}
\submitjournal{ApJL}
\shorttitle{Quiescent galaxies at $z \gtrsim 2.5$: observations vs. models}
\shortauthors{Cecchi et al.}

\begin{document} 

   \title{Quiescent galaxies at $z \gtrsim 2.5$: observations vs. models}

\correspondingauthor{Rachele Cecchi}
\email{rachele.cecchi@inaf.it}

\author{Rachele Cecchi}
\affil{Universit\`a di Trieste, Dipartimento di Fisica, via Valerio 2, I-34127 Trieste, Italy}
\affiliation{INAF -- Osservatorio Astronomico di Trieste, via G.B. Tiepolo 11, I-34143 Trieste, Italy} 
\affiliation{Universit\`a di Bologna, Dipartimento di Fisica e Astronomia, Via P. Gobetti 93/2, I-40129, Bologna, Italy}

\author{Micol Bolzonella}
\affiliation{INAF -- Osservatorio di Astrofisica e Scienza dello Spazio di Bologna, via P. Gobetti 93/3, I-40129 Bologna, Italy}

\author{Andrea Cimatti}
\affiliation{Universit\`a di Bologna, Dipartimento di Fisica e Astronomia, Via P. Gobetti 93/2, I-40129, Bologna, Italy}
\affiliation{INAF -- Osservatorio Astrofisico di Arcetri, Largo Enrico Fermi 5, I-50125 Firenze, Italy}

\author{Giacomo Girelli}
\affiliation{INAF -- Osservatorio di Astrofisica e Scienza dello Spazio di Bologna, via P. Gobetti 93/3, I-40129 Bologna, Italy}
\affiliation{Universit\`a di Bologna, Dipartimento di Fisica e Astronomia, Via P. Gobetti 93/2, I-40129, Bologna, Italy}
 
\begin{abstract}
The presence of massive quiescent galaxies at high redshifts is still a challenge for most models of galaxy formation. The aim of this work is to compare the observed number density and properties of these galaxies with the predictions of state-of-the-art models.

The sample of massive quiescent galaxies has been selected from the COSMOS2015 photometric catalogue with $z_{\rm phot}\geq 2.5$, $\log (M_*/M_\odot)\geq 10.5$ and $\log(\mathrm{sSFR\,[yr^{-1}]})\le -11$. The photometric SEDs of the selected galaxies have been thoroughly analyzed based on different stellar population synthesis models. The final sample includes only those galaxies qualified as quiescent in all SED fitting runs.

The observed properties have been compared to theoretical models: the number density of quiescent galaxies with $10.5 \leq \log(M_*/M_\odot) < 10.8$ is reproduced by some models, although there is a large scatter in their predictions. Instead, very massive $\log(M_{*}/M_{\odot}) \geq 10.8$ are underpredicted by most of the current models of galaxy formation: some of them, built on the CARNage simulation, are consistent with data up to $z \sim 4$, while at higher redshifts the volume of the considered simulation is too small to find such rare objects.
Simulated galaxies which match the observed properties in the $\mathrm{sSFR}-M_*$ plane at $z\sim 3$ have been analyzed by reconstructing their evolutionary paths: their merger trees suggest that AGN feedback could be the key process allowing for a rapid quenching of the star formation at $z\gtrsim 4$ and that its treatment should be improved in models.

\end{abstract}

   \keywords{Galaxies: evolution --- Galaxies: formation --- Galaxies: high-redshift --- Galaxies: star formation}

\section{Introduction}
\label{introduction}

The discovery of massive quiescent galaxies at cosmological distances \citep{Cimatti04} opened a new possibility to investigate the physical processes leading to formation and assembly of the present-day E/S0 systems. Further observations at optical and near-infrared wavelengths unveiled a substantial population of high redshift (up to $z\gtrsim 3$) and massive galaxies (stellar mass $M_{*} \sim 10^{11}M_{\odot}$) with weak or absent star formation \citep[hereafter quiescent galaxies, or QGs; e.g.][]{Wiklind08, Mancini09, Capak11, Gobat12, Caputi12, Caputi15, Huang14, Nayyeri14, Glazebrook17, Schreiber18, Merlin18, Osejo19}. However, the existence of such population in the early universe is still a challenge for models of galaxy formation; in fact, these objects should have experienced intense episodes of star formation with a large amount of stellar mass formed at even higher redshifts, followed by the aging of the stellar population that appears already evolved at $z > 3$. This implies that their ages are consistent with being almost as old as the universe at their redshift. Investigating the nature and evolutionary paths of quiescent galaxies, and especially the most massive ones, is essential to understand the baryonic processes and the mass assembly leading to the formation of these rare galaxies \citep[e.g.][]{Somerville15, Naab17}.

Quiescent galaxies at high redshift are very red at optical observed wavelengths, due to their old stellar population and $k$-correction. The published samples are all photometrically selected, and their study is based on the SED fitting analysis \citep{Merlin18, Nayyeri14, Mawatari16, Straatman14, Wiklind08}; secure spectroscopic identification of these systems is still very limited \citep{Gobat12,Glazebrook17,Schreiber18}, and will greatly benefit of the advent of JWST.

A variety of semi-analytic models (hereafter SAMs) have been used in order to understand how and when these galaxies formed and how they quenched their star formation at high redshift \citep[e.g.][]{Henriques15,Guo11, Guo13, Naab14, Hirschmann16, Fontanot17b}.

The aim of this work is to select a sample of reliable QGs at high redshifts and to compare their number density and properties with state-of-the-art predictions of galaxy formation models. To this purpose, particular attention was paid to the optimal selection of QGs based on photometric redshifts and SED fitting in order to maximize the purity of the final sample.

\section{Selecting quiescent galaxies at high redshift}
\label{selection}

\begin{figure*}
    \centering
    \includegraphics[width=0.99 \textwidth]{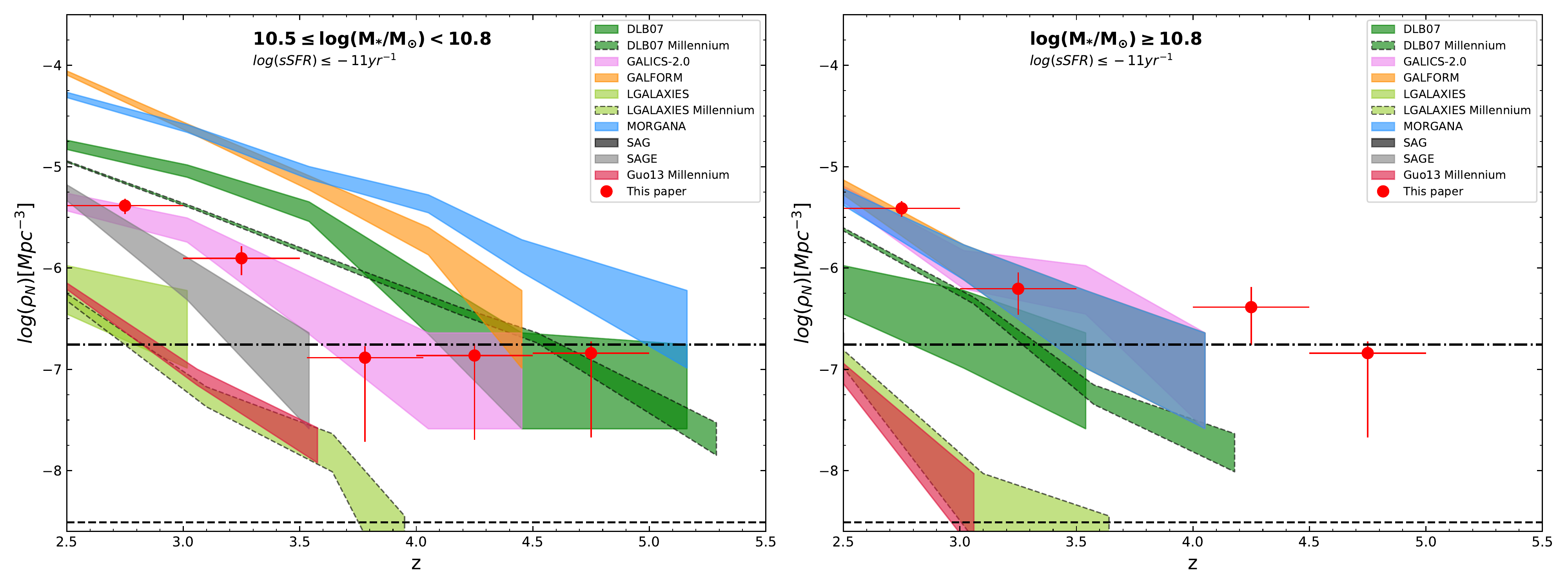}
    \caption{Number densities of QGs in two stellar mass regimes. Red dots represent observed galaxies selected in Sec.~\ref{sedfit} from COSMOS with their Poissonian error bars. Colored shaded regions represent $\pm 1\sigma$ data from CARNage and Millennium simulations (same colors as CARNage when the same model is present, but with black dashed lines). The dashed and dash-dotted horizontal black lines represent the number densities expected for 1 object in the simulation volume of Millennium and CARNage respectively.}
    \label{fig:num_dens}
\end{figure*}

We used the dataset COSMOS2015 \citep{Laigle16} containing multi-waveband deep photometric observations, photometric redshifts and physical properties for $1\,182\,108$ objects extracted from the $\sim 2 \, \mathrm{deg}^2$ area of the COSMOS survey.

For the first selection, we considered the region of $1.38 \, \mathrm{deg}^2$ with best photometric quality, reducing the sample to $536\,077$ objects, and we focus our study on galaxies at $z_{\rm phot} \equiv {\rm ZPDF} \geq 2.5$, where ${\rm ZPDF}$ is the median of the likelihood distribution.  Quiescent galaxies are selected by means of the parameter ${\rm CLASS}=0$, reflecting the classification in the rest-frame color-color plane ${\rm NUV}-r$ vs $r-J$ \citep{Ilbert10, Ilbert13}.
We applied a selection in magnitudes, $m_{[4.5 \mu m]}\leq 24$: this  threshold ensures an accurate photometry ($\mathrm{S/N}>3$ for $m_{[4.5 \mu m]}\leq 24$), and the chosen NIR waveband represents a proxy of galaxies stellar masses also at high redshift. This selection translates into a limit of stellar masses $\geq 10^{10.2} M_{\odot}$ for synthetic models with exponentially declining SFH with characteristic time $\tau \le 0.3\,{\rm Gyr}$, Chabrier IMF, and formation redshift $z_{\rm form}\sim 6$, generally used to reproduce the SEDs of QGs. 
Finally, we excluded objects with a significant emission at $24 \mathrm{\mu} m$ to mitigate a degeneracy with dusty star-forming galaxies, and X-rays sources.  We obtained a sample of $376$ candidate massive QGs at $z_{\rm phot} \geq 2.5$.

\subsection{Refining the sample of quiescent galaxies}
\label{sedfit}

The photometric SEDs of the selected galaxies have been analyzed to refine the sample and maximize its reliability.

In order to take into account the possible degeneracies and uncertainties in the photometric redshift estimate, we analyzed their probability distribution functions (PDF), provided with the COSMOS2015 catalog. We visually inspected the PDF$(z)$ of each object to check the  existence of  anomalous or multi-modal distribution which would make the photometric redshift estimate uncertain. We classified the objects in three classes: \textit{Type 0} ($325$ objects) with Gaussian-like probability distributions and a clear peak around the value ZPDF; \textit{Type 1} ($41$ objects) have a wide and asymmetric PDF$(z)$; \textit{Type 2} ($10$ objects) have a multi-peaked distribution. For \textit{Type 1} and \textit{2} objects we performed the SED fitting also at the peak(s) of the PDF$(z)$, as the latter are significantly different from ZPDF, to verify if the choice of one of the two changes the physical properties and the nature of selected galaxies.

We performed a new SED fitting, optimized both to consider a wide range of parameters and to reduce possible degeneracies. We used broad-band magnitudes from $u$ to $4.5 \rm \mu m$ bands, and followed Eqs.~9 and 10 of \citealt{Laigle16} to obtain pseudo-total magnitudes. 
To this aim we used the code \textit{hyperzmass} \citep{Bolzonella00,Bolzonella10} to find the best fit SED, fixing the redshift at one of the photometric redshifts indicated by the PDF$(z)$.  

We included in the final sample only galaxies that are quiescent in all $4$ runs of SED fitting described in Appendix~\ref{appendix_sedfit}, adopting a cut in the specific star formation rate $\log(\mathrm{sSFR\,[yr^{-1}]}) \leq -11$ as the definition of quiescent galaxy.  Our final sample contains $100$ QGs with $z \geq 2.5$ and $M_* \geq 10^{10.2} M_{\odot}$.


\section{Observed number densities and comparison with models}
\label{numdens}

We computed the number densities of our observed sample of galaxies after applying a further cut in stellar mass, $\log(M_{*}/M_{\odot}) \geq 10.5$, corresponding to the completeness mass estimated at $z \sim 3$ for our sample, following the method by \cite{Pozzetti10}. A similar value has been obtained by \cite{Laigle16} for a slightly different selection. This choice ensures that our sample is both pure, as discussed in Sect.~\ref{sedfit}, and complete. 

The sample was splitted into two stellar mass bins to explore the possibility of different quenching mechanisms (e.g. \citealt{Peng10}): \textit{intermediate masses} galaxies, with $10.5 \leq \mathrm{log}(M_{*}/M_{\odot}) < 10.8$ ($N=47$), and \textit{massive} galaxies with $\mathrm{log}(M_{*}/M_{\odot}) \geq 10.8$ ($N=41$).

We compared the number density of our best candidate QGs sample with the predictions by different SAMs. We made use of the CARNage dataset \citep{Knebe18,Asquith18}, a collection of different models of galaxy formation run on the same cold dark matter simulation in a box of comoving size $125\, h^{-1}\, \mathrm{Mpc}$. The $7$ different SAMs used in this work are: DLB07 \citep{Delucia07}, GALFORM \citep{Gonzalez14}, GalICS-2.0 \citep{Cattaneo17}, LGALAXIES \citep{Henriques13}, Morgana \citep{Monaco07}, SAG \citep{Cora18} and SAGE \citep{Croton16}. For a detailed description see \cite{Knebe15} and references therein.

We also computed the number densities obtained by SAMs applied to the Millennium simulation: DLB07 \citep{Delucia07}, LGALAXIES \citep{Henriques13} and Guo \citep{Guo13}. The advantage of the Millennium simulation is the larger volume (having a box of comoving side $\sim 500\, h^{-1}\, \mathrm{Mpc}$) that minimizes the uncertainties due to cosmic variance and is more suitable for rare objects such as high-$z$ QGs. Since DLB07 and LGALAXIES models are also available in the CARNage simulation, we verified the consistency of the two datasets, using the so-called ``uncalibrated'' version in CARNage\footnote{The CARNage dataset is based on different mergertrees and different definitions and estimates of physical properties, for this reason it needs to be recalibrated to be consistent with other simulations. See \cite{Knebe18} for a more detailed description about calibration.} In the following, for the CARNage simulation we used data with their final calibration (``calibration 2'', see \citealt{Knebe18} for details).

To select QGs from simulations we imposed the same thresholds in sSFR and $M_{*}$, after convolving the stellar masses with a scatter of $0.04(1+z) \, \mathrm{dex}$, where $z$ is the redshift of each galaxy, to reproduce the observational errors of COSMOS on simulated stellar masses \citep{Ilbert13}. The number densities without the convolution in stellar masses are shown in Fig.~\ref{fig:num_dens_noEB} of Appendix.~\ref{appendix_edd}. When necessary, we converted stellar masses to the Chabrier IMF using Eq.~2 of \cite{Knebe15}. 
We analyzed the snapshots at $z = 2.5, 3, 3.5, 4, 4.5, 5$ in order to match the redshift coverage of our sample.

For stellar masses in the range of $10.5 \leq\mathrm{log}(M_{*}/M_{\odot}) < 10.8$ (Fig.~\ref{fig:num_dens}, left panel), some SAMs underestimate systematically the number of these objects at all redshifts (i.e. Guo13, LGALAXIES and SAGE), while others overestimate it (i.e. Morgana and GALFORM). The SAMs that better reproduce the observations in this regime are DLB07 and GALICS-2.0. The agreement between models and observations becomes drastically worse at $z \gtrsim 4$ where most models do not predict the existence of these objects. This disagreement can be due to a real absence of this type of objects at such a high redshift, or to an insufficient volume, being the latter particularly relevant in the case of CARNage, as shown in the plot by the horizontal limit identifying the number density for a single object contained in the simulation box.
The differences between the predictions of the same SAM present in both CARNage and Millennium datasets depend on the different calibrations adopted by the authors in the two simulations.

In the highest mass regime (Fig.~\ref{fig:num_dens}, right panel), three SAMs (GALFORM, Morgana and GalICS-2.0) applied to CARNage are in good agreement with the number densities of the observational sample up to $z\lesssim 4$. At higher redshifts the volume of the CARNage simulation hampers any comparison with models: the lower limit of the number densities is marked by the horizontal dot-dashed line in the plot, showing that at $z>4$ the volume is too small to find such rare objects.


\section{How to make a galaxy quiescent at high redshift}
\label{mergertrees}

\begin{figure}
\includegraphics[width=0.49\textwidth]{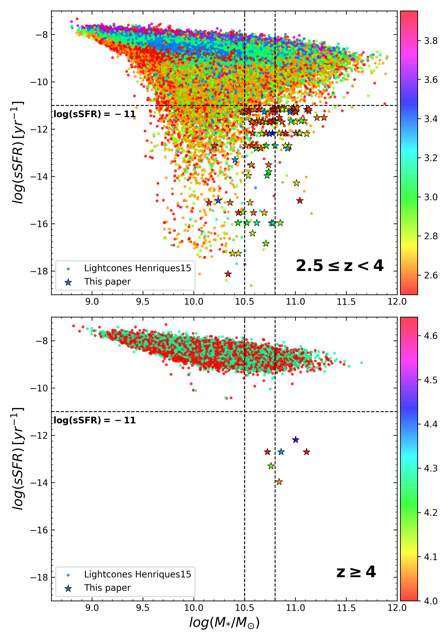}
  \caption{sSFR vs $M_{*}$ diagram in two redshift ranges ($2.5 \leq z<4.0$ in the top panel and $z \geq 4$ in the bottom one) for the observed galaxies selected in this work (stars with black outlines) and for the simulated galaxies with $m_{[4.5\mu m]}\le 24$ selected from the $24$ lightcones of \citet{Henriques15} in the Millennium database (small points). All points are colored according to their redshift, as shown by the two colorbars on the right. The horizontal black lines show the threshold in $\log(\mathrm{sSFR})$ for the quiescent selection, while vertical lines delimit the two mass regimes.}
\label{fig:ssfr_M_tot}
\end{figure}

\begin{figure*}
    \centering
    \includegraphics[width=0.9\textwidth]{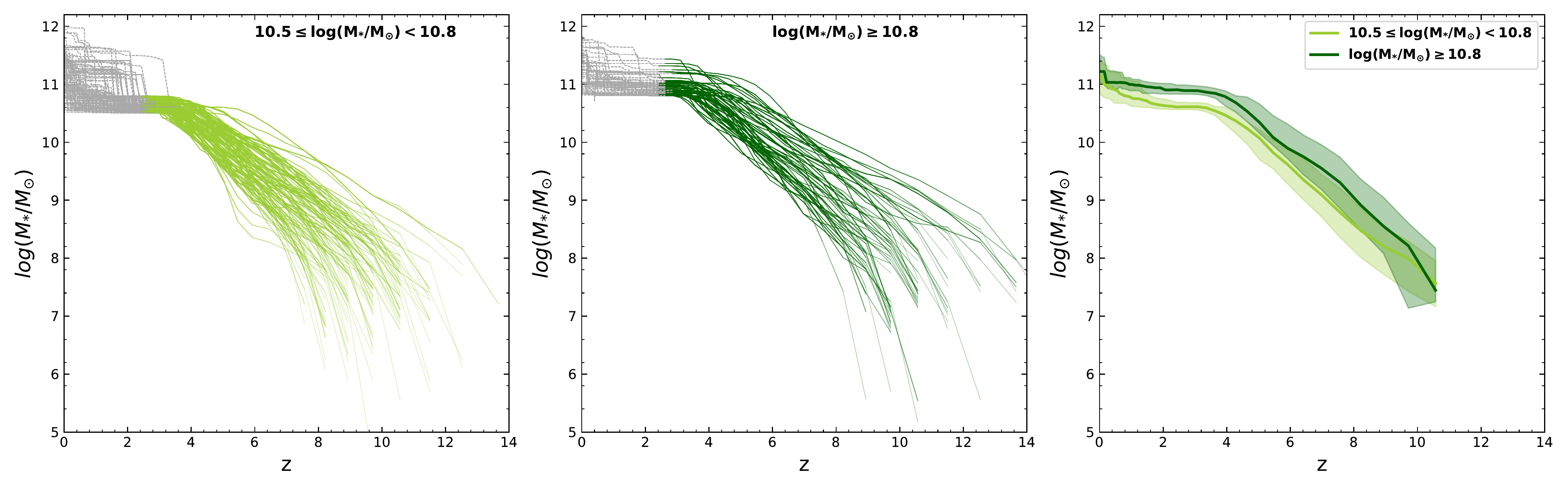}
    \includegraphics[width=0.9\textwidth]{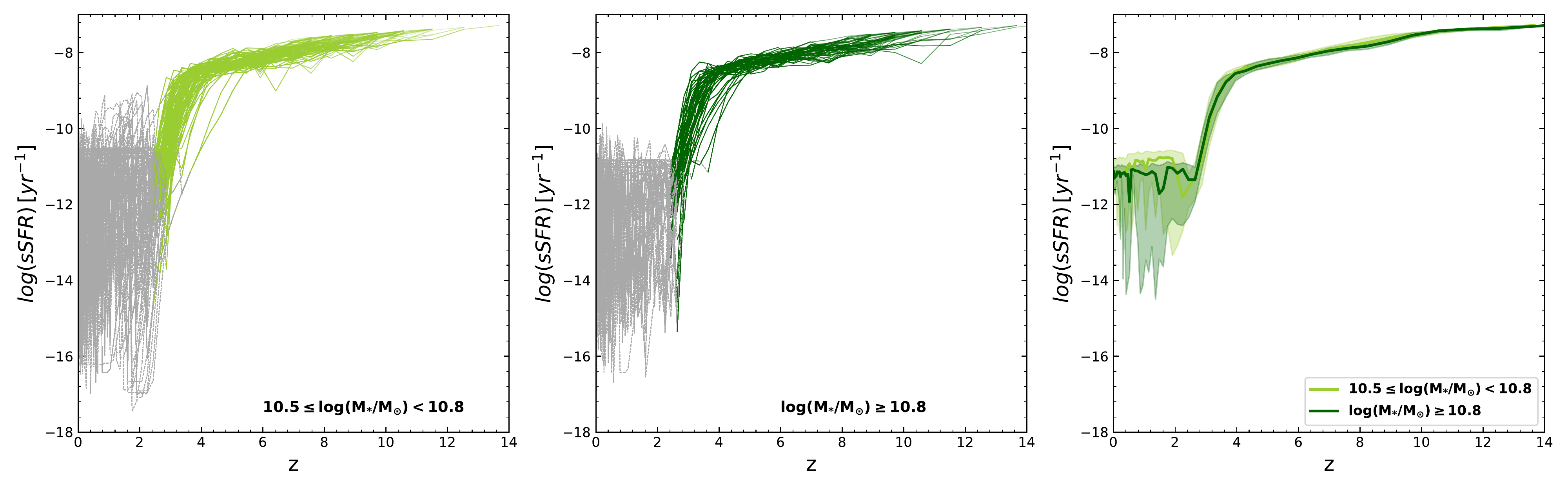}
    \caption{Evolution of the stellar mass (top) and sSFR (bottom) with redshift for simulated QGs selected in the SAM LGALAXIES from the lightcones in the Millennium database. Left panels: galaxies with $10.5 \leq \log(M_{*}/M_{\odot}) < 10.8$ (light green for progenitors, gray for descendants). Central panels: galaxies with $\log(M_{*}/M_{\odot}) \geq 10.8$ (dark green for progenitors, gray for descendants). Right panels: median and quartiles of the distributions in the first two plots.} 
    \label{fig:trees_henriques}
\end{figure*}

In this section, we attempt to reconstruct the evolutionary paths of the observed QGs making use of the information available in the models. This kind of analysis can be performed by using the merger trees, i.e. the chronological history of physical properties obtained by identifying descendants and progenitors, of galaxies extracted from the models with the same criteria used to select the real galaxies of our sample. 

In Fig.~\ref{fig:ssfr_M_tot} we show the sSFR vs $M_{*}$ diagram for observed galaxies selected in Sect.~\ref{sedfit} and simulated galaxies of \cite{Henriques15} as extracted from the Millennium database. For the latter, we used the combination of the $24$ available lightcones (of area $3.14\, \mathrm{deg}^2$ each, for a total area $\sim 50$ times bigger than the observed one, with a magnitude selection $m_{[4.5\mu m]}\le 24$ using BC03 SPS models) instead of the snapshots, to perform a selection that includes the redshift and apparent magnitude distributions, more similar to observations. From this plot is is clear that the study of the merger trees can only be done selecting QGs from models at $z\la 4$, where a fraction of objects has sSFR and $M_{*}$ compatible with the observed ones, while at higher redshift no massive QGs are present in the models. In the following, we focus only on simulated galaxies with similar sSFR and $M_{*}$ to the observed ones, assuming that sharing similar sSFR and $M_{*}$ means having comparable history of formation and evolution.

\subsection{The growth of stellar mass}
\label{massgrowth}

We built the merger trees of some of the SAMs from the CARNage dataset that best reproduce the number densities of massive galaxies at high redshift (DLB07, Morgana and GALFORM), selecting galaxies in the snapshot at $z=3$, while for LGALAXIES we used the lightcones and galaxies selected in the same redshift range of observed ones. In both cases the galaxies were selected after convolving their stellar masses with the observational error (Eddington bias) as in Sect.~\ref{numdens}, while their evolution is analyzed using the intrinsic properties provided by each SAM.

Figure~\ref{fig:trees_henriques} (top panels) shows the evolution of $M_*$ with redshift for QGs selected in the SAM LGALAXIES in the redshift range $2.5 \leq z \leq 4$ for the two stellar mass regimes. The stellar mass of the simulated QGs increases as the redshift decreases, reaches a sort of plateau around $z \sim 3$ and then grows again at lower redshift, at a rate depending on the selection mass. The plateau corresponds to an interruption of the stellar formation within these galaxies, already quiescent at these redshift. There are no significant differences in mass growth trends for the two sample of intermediate and massive galaxies, suggesting a similar evolution at least until reaching the plateau. However, at lower redshift the descendants of individual intermediate mass galaxies experience a discontinuous mass growth, with sudden growths signatures of merging phenomena. At $z=0$ the two galaxy samples reach more or less the same median mass. 

We also analyzed the average evolution of the specific star formation rate for LGALAXIES (Fig.~\ref{fig:trees_henriques}, bottom panels). All the QGs selected in LGALAXIES at $z\sim 3$ underwent a steep decrease of their star formation between $z \sim 4$ and $z \sim 3$. Subsequently, the sSFR of massive galaxies remains low, while intermediate mass ones experienced sporadic and sudden increase of the sSFR, suggestive of merging events with gas rich galaxies. Even so, the median trend is very similar in the two mass regimes. 

\begin{figure*}
    \centering
    \includegraphics[width=0.9\textwidth]{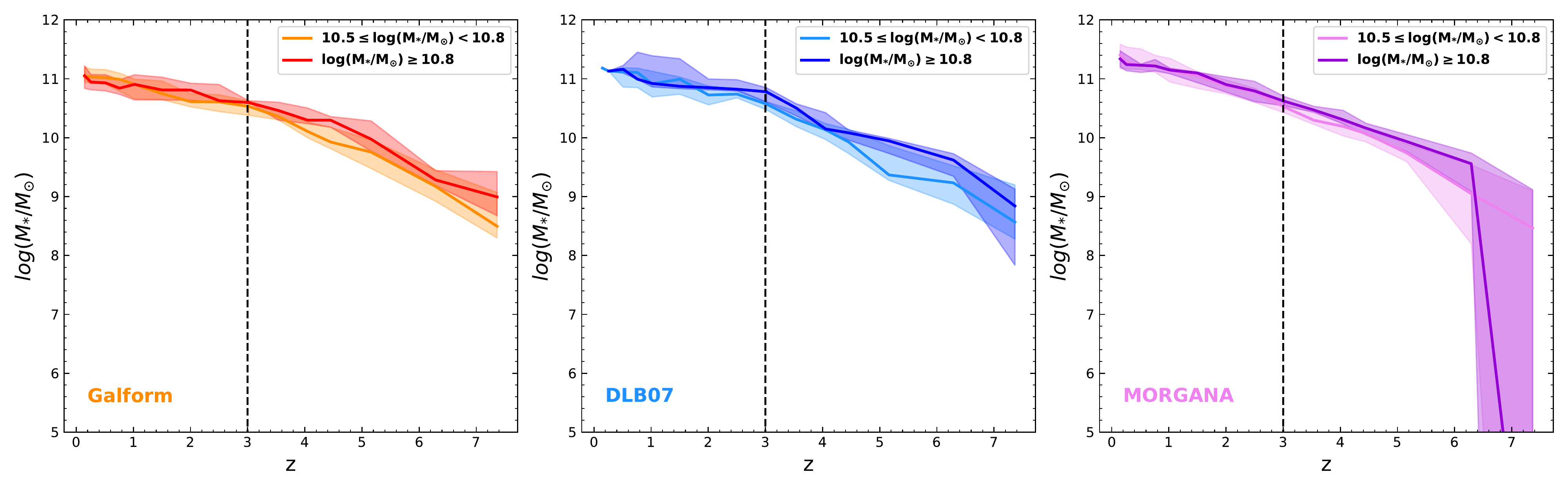}
    \includegraphics[width=0.9\textwidth]{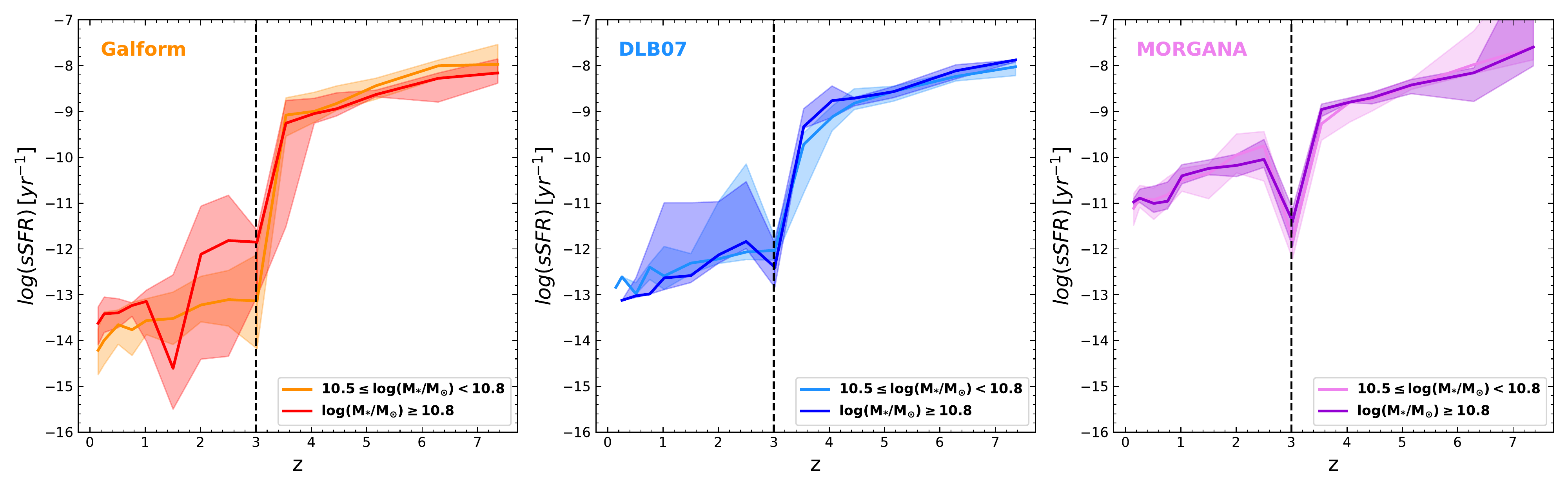}
    \caption{Evolution of the median stellar masses (top) and sSFR (bottom) for simulated QGs in the two mass ranges selected in the SAMs Galform (left), DLB07 (central) and Morgana (right) from CARNage database. The light- and dark-coloured lines refer to galaxies belonging to intermediate and massive populations respectively. Shaded areas represent the quartiles of the distribution, while the black dotted vertical line represents the redshift of the box used for selection.} 
    \label{fig:trees_carnage}
\end{figure*}

To better understand the low redshift evolution ($z<3$, after the observed plateau) of these galaxies, we remind that a reactivation of the star formation at these redshifts can be due to two main processes: merging between a gaseous satellite and a central galaxy that cause a sudden starburst, or gas cooling processes that can occur in galaxies of these masses. We deduce that a sudden jump in mass corresponds to a merging phenomenon, while a more gradual growth is due to gas cooling intrinsic to the galaxy itself. The combination of these phenomena produces the median trend observed at low redshift in Fig.~\ref{fig:trees_henriques}.

As mentioned above, the data from the CARNage simulation are only available at fixed snapshots: we therefore selected galaxies in the snapshot at $z=3$, because it is the most common redshift of our observed sample, and we followed their progenitors up to the highest redshift available in CARNage snapshots, i.e. $z \sim 7$. 
Figure~\ref{fig:trees_carnage} (top panels) shows the evolution of the median stellar mass of progenitors and descendants of the selected QGs at $z=3$ in Galform, DLB07, and Morgana SAMs. As seen for the number density, most massive galaxies are very rare in these SAMs and therefore their statistical description is more uncertain. Nonetheless, some trend can be noticed: in the three considered models there are no significant differences in the median evolution of the mass growth for intermediate and massive QGs. A steadier increase of stellar mass is visible, although with a different rate, making them slightly different at $z=0$. In Galform and DLB07 it is visible a variation of the slope of mass growth from progenitors to descendants, with the mass increase slowing down at $z<3$, a clue of the parallel decrease of the star formation activity.
A plateau at redshift $z\la 3$, indication of an interruption of the SFR like in LGALAXIES, is only visible for massive QGs in DLB07 model. 

In the bottom panels of Fig.~\ref{fig:trees_carnage} we show the median evolutionary paths of the sSFR for the SAMs. They present a variety of trends: 
galaxies in Galform and DLB07 models are characterized by a quick drop of the sSFR at $3<z<4$, as seen also in LGALAXIES, followed by a continuous decline; QGs selected in Morgana instead experience a constant decrease of the sSFR from the formation to the present epoch, maintaining a higher level of star formation than the other two models.
The evolution of the sSFR in Morgana shows a sharp dip at the redshift of selection, and the same behavior is present in the other models, although less prominent. The reason is the rapid activation and suppression of the SFR occurring in timescales shorter than the interval between the snapshots, depending on the implemented mechanism of gas accretion. The net result is a sSFR that fluctuates, so that galaxies selected as quiescent at $z \sim 3$ can have an active star formation at lower redshifts.  These SAMs (especially Morgana) are therefore able to reproduce the observed number densities thanks to the rapid fluctuations of SFR, but as a consequence they would have colors different from those of observed galaxies, and they would not agree with our selection of quiescent galaxies which is based on their SEDs.

\subsection{The role of massive black holes}
\label{bh}

\begin{figure*}
    \centering
    \includegraphics[width=0.9\textwidth]{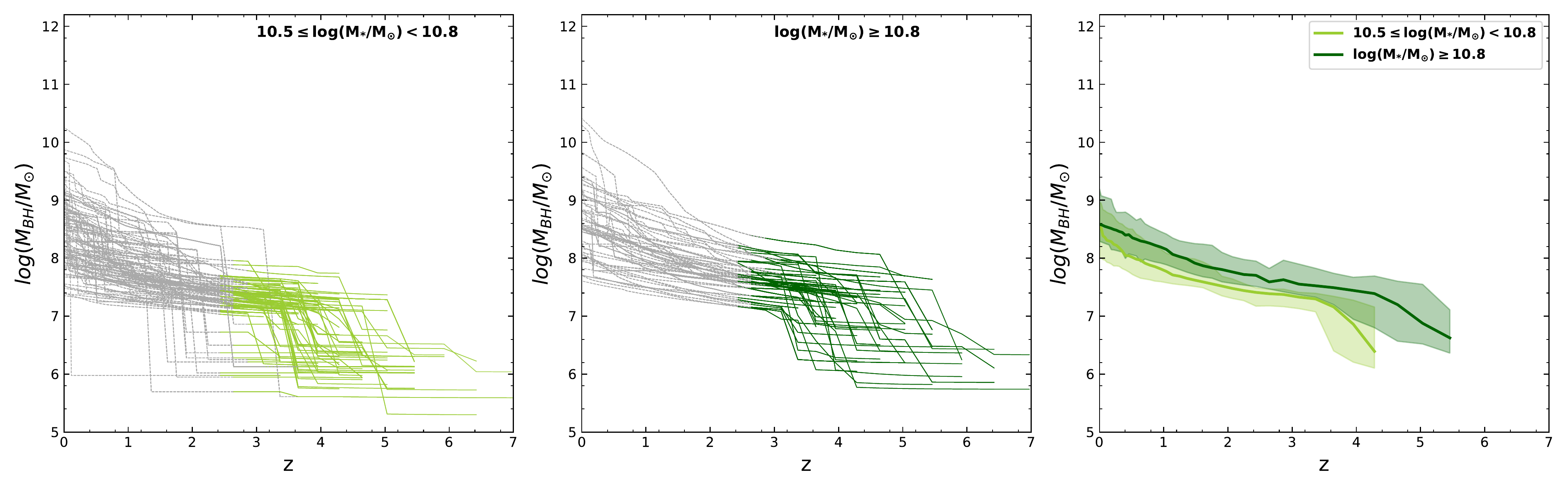}
    \includegraphics[width=0.9\textwidth]{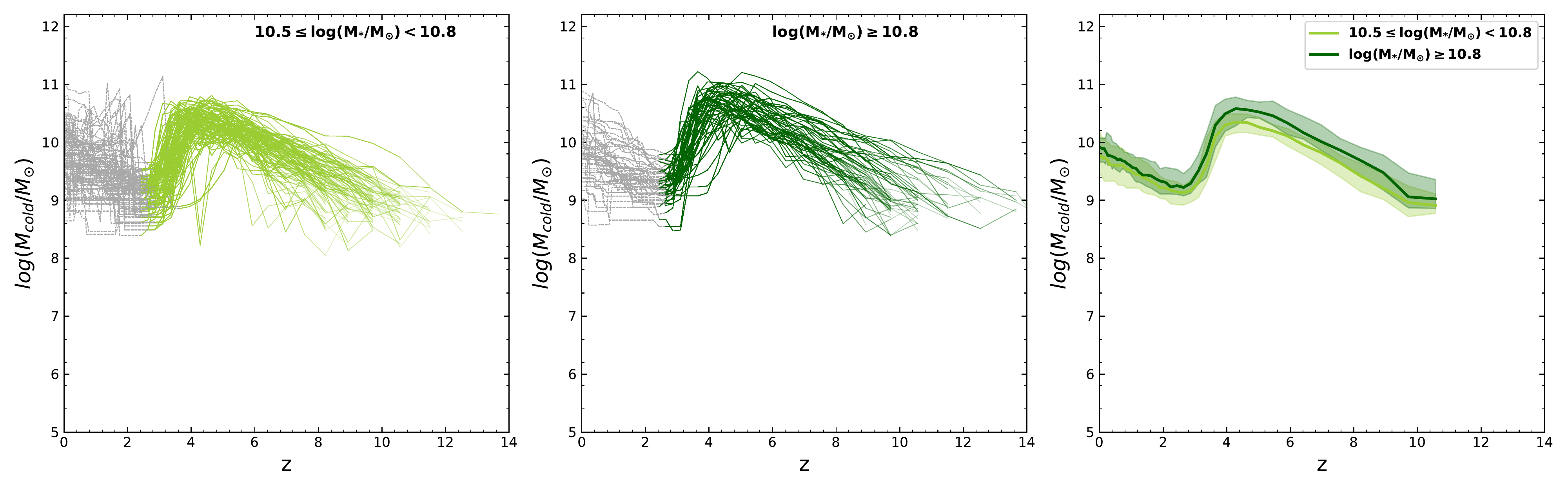}
    \caption{Trend of the black hole mass (top) and cold gas mass (bottom) with redshift for simulated QGs selected in the SAM LGALAXIES from the lightcones in the Millennium database. Colors and symbols as in Fig.~\ref{fig:trees_henriques}.} 
    \label{fig:trees_mbh_mcold_henriques}
\end{figure*}

\begin{figure*}
    \centering
    \includegraphics[width=0.9\textwidth]{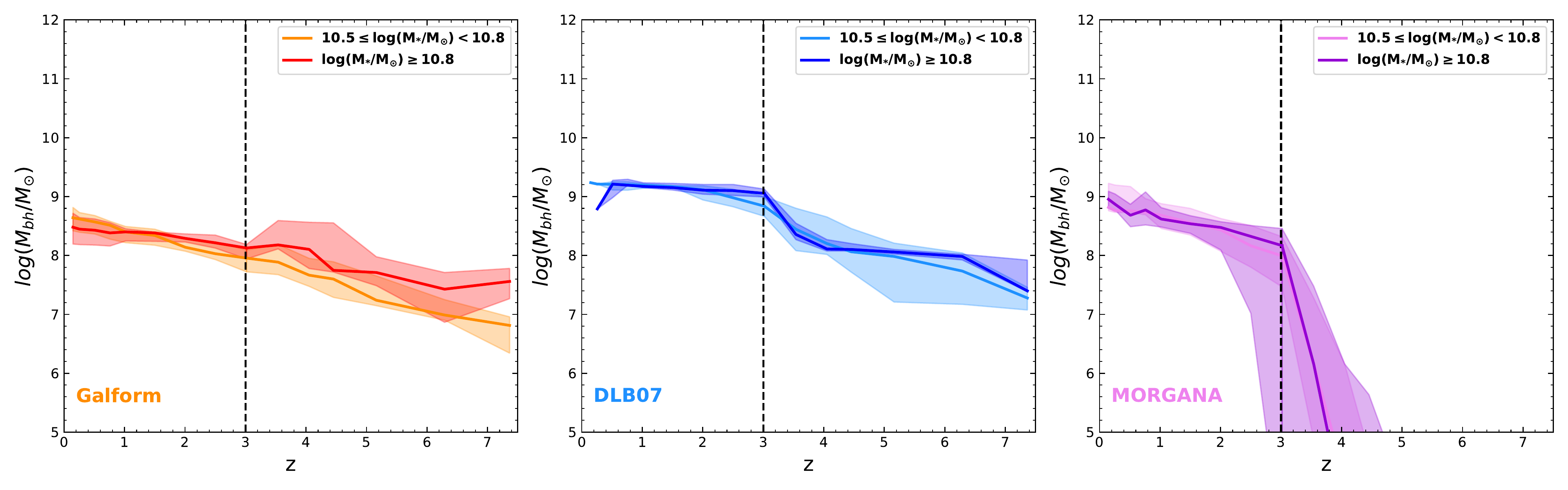}
    \includegraphics[width=0.9\textwidth]{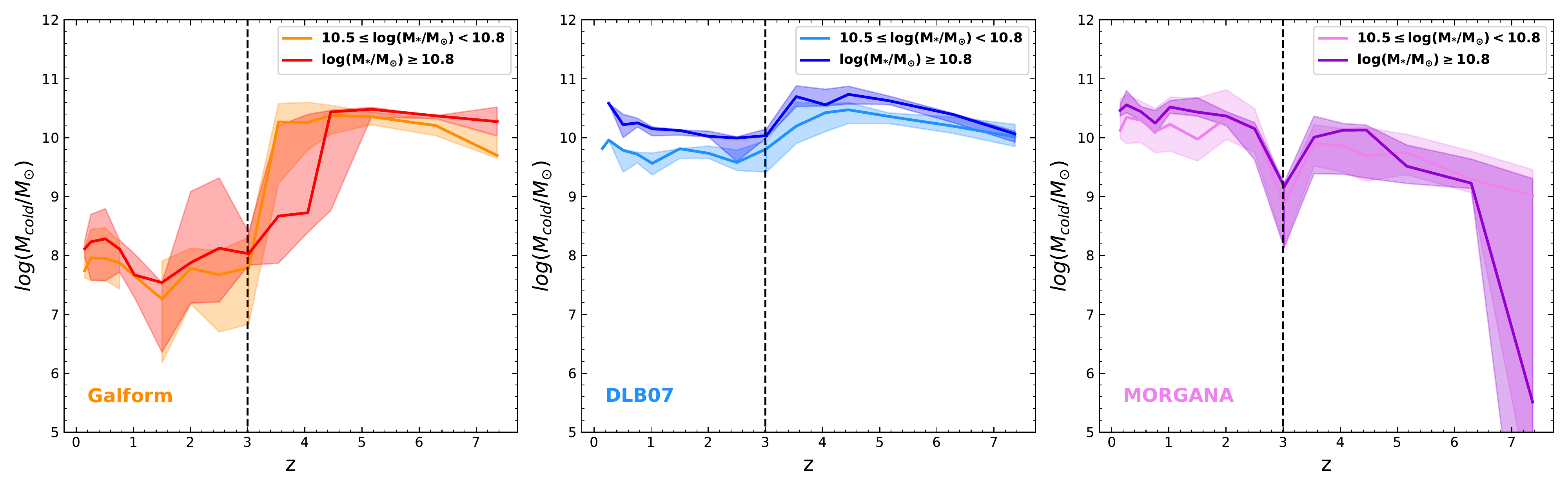}
    \caption{Evolution of the median black hole mass (top) and cold gas mass (bottom) for simulated QGs in the two mass ranges selected in the SAMs Galform (left), DLB07 (central) and Morgana (right) from CARNage database. Colors and symbols as in Fig.~\ref{fig:trees_carnage}.} 
    \label{fig:trees_carnage_mbh_mcold}
\end{figure*}

To better understand the physical mechanisms that produce the trends described in Sect.~\ref{massgrowth} and lead to the formation of QGs at high redshift, we analyzed the possible significant events that heated or removed the gas hence halting the star formation \citep{Peng2015,Cresci2018,Man2018}, and their implementation in the considered SAMs, taking into account that at the stellar masses and redshifts considered in the present work the environmental quenching is not expected to play a major role.

The negative AGN feedback is commonly described by the contribution of two main processes: quasar mode, with an efficient growth of $M_{\rm BH}$, in general triggered by mergers with accretion of cold gas; radio mode, with a slow and long-lived growth of $M_{\rm BH}$ that heats the gas in the halo, common in massive galaxies. Moreover, stellar winds produced by SNe can also quickly remove the cold gas, making it unavailable for star formation.
In the considered SAMs, the recipes accounting for the radio mode are mostly derived from \citet{Croton06}, with different efficiencies as a function of redshift. The radio mode in these models is providing the feedback necessary to keep galaxies passive at low redshifts. The quasar mode feedback, instead, is effectively implemented only in Morgana, while in the other SAMs it is only responsible for the BH mass growth through the cold mass accretion, but it does not affect directly the SFR.

To understand the relevance of these processes in the considered models we analyzed the evolution of the mass of the central black hole $M_{\rm BH}$ and the mass of cold gas $M_{\rm cold}$.
Their evolution for LGALAXIES model is illustrated in Fig.~\ref{fig:trees_mbh_mcold_henriques}: the sudden drop observed in the sSFR corresponds to a sharp decrease of $M_{\rm cold}$ at $z\la 4$. At lower redshifts a mild increase of $M_{\rm cold}$ is observed, possibly a result of wet mergers and gas cooling phenomena as described in Sect.~\ref{massgrowth}, especially for intermediate mass galaxies, mirrored by the reactivation of the star formation observed in Fig.~\ref{fig:trees_henriques}. At high redshifts, $M_{\rm BH}$ appears to evolve stepwise, again symptom of merging: although the quasar mode feedback is not implemented in this model, the consumption of the cold gas due to collisional starburst and accretion to the BH triggered by mergers is the responsible of quenching the star formation. 

In Fig.~\ref{fig:trees_carnage_mbh_mcold} we show the median evolution of 
$M_{\rm BH}$ and $M_{\rm cold}$ for the three models from CARNage. Very different trends are observed: while the growth of $M_{\rm BH}$ in GALFORM and DLB07 is somehow similar to the LGALAXIES one, with an overall gradual increase, possibly with some hint of a fast growth at $3\la z\la 4$ in DLB07, in Morgana the BHs abruptly appears at $z\sim 4$. 
The growth of the BHs is reflected in the trend of $M_{\rm cold}$, that correspondingly decrease in GALFORM and DLB07, but at a very different rate. Again, Morgana foresees a very different trend, with a decrease of the cold gas mass only at the selection redshift as a consequence of the BH appearance, but with a subsequent growth of descendants' $M_{\rm cold}$. 

Taken together all the trends, we derive the hint that the physical processes triggered by mergers, with the accretion of cold gas to the BH and violent star formation induced by collision, is the main responsible of the formation of QGs at high redshift. However, its current implementation in SAMs is not fully reproducing the number of observed galaxies. 


\section{Conclusions}
\label{conclusions}

We selected a complete and well-controlled sample of QGs at $z\ge 2.5$ defined as galaxies with $\log (M_*/M_\odot) \ge 10.5$ and ${\rm sSFR} \le 10^{-11}\,{\rm yr}^{-1}$. Their properties were compared with those predicted by different SAMs. At $z \la 4$ and $10.5\le \log(M_*/M_\odot) < 10.8$, the models predict a very wide range of number densities, and some of them agree with the observations.

At $z>4$ and $\log (M_*/M_\odot) \ge 10.8$, most models fail to reproduce the observations because such galaxies are not predicted at all. Three SAMs are instead in good agreement with the observed number densities up to the limits of the simulation volume. The merger trees of the simulated galaxies which reproduce the observed ${\rm sSFR}$ and $M_*$ show that quiescence is achieved mostly through merging and/or BH growth. For some models the agreement on the number density is the result of rapid fluctuations of the SFR, that would produce SEDs inconsistent with the ones of the observed galaxies of the sample selected in the present study.

The process of star formation quenching at high redshifts needs to be further investigated in order to explain the existence of such massive and mature systems at early cosmic epochs.

\begin{acknowledgements}
We acknowledge the grants PRIN MIUR 2015 and ASI n.2018-23-HH.0, Fabio Fontanot, Gabriella De Lucia, and Pierluigi Monaco for enlightening discussion, Alexander Knebe and Bruno Henriques for their help in using the models, and the anonymous referee for the valuable suggestions.  

\end{acknowledgements}

\bibliographystyle{aasjournal} 
\bibliography{rachele}

\begin{appendix}
\section{SED fitting parameters}
\label{appendix_sedfit}

We used two different stellar population synthesis (SPS) models: \citet{B&C03} and \citet{Maraston05} (hereafter BC03 and M05), whose difference is mainly due to a more significant contribution of TP-AGB stars in M05 models. We also used different star formation histories (SFH), i.e. exponentially declining ($\mathrm{SFR} \propto \tau ^{-1}e^{-t/\tau}$) and delayed SFH ($\mathrm{SFR} \propto \tau ^{-2} t e^{-t/\tau}$) for QGs, both with three different characteristic timescales $\tau$ ($0.1$\,Gyr, $0.3$\,Gyr, $0.6$\,Gyr), and a model with constant star formation to rule out contamination from star-forming galaxies. We also used two different values of metallicity: solar ($Z=0.02$) and sub-solar ($0.4 Z_{\odot}$ for BC03 and $0.5 Z_{\odot}$ for M05). The two SPS models assume different IMFs: Chabrier for BC03 and Kroupa for M05. We used \citet{Calzetti00} as extinction law with a maximum extinction of $A_{V}=5$, a much larger limit than the one adopted in the work by \citet{Laigle16}, to allow the fit also with very dusty star-forming galaxies. 

Varying all the parameters just described we obtained $32$ templates divided into $4$ runs of SED fitting, described in Table~\ref{tab:list}. 

In all the runs we adopted a $\Lambda$CDM cosmology with $H_{0}= 70 \, {\rm km \,s^{-1}\,Mpc^{-1}}$, $\Omega_{\rm m}=0.3$ and $\Omega_{\Lambda}=0.7$. 

\begin{table}
\caption{Summary of the 4 runs of the SED fitting analysis used in this work.} 
\label{tab:list}
\centering
\begin{tabular}{c|cccc} 
\hline\hline
\textbf{Run} & \textbf{1} & \textbf{2} & \textbf{3} & \textbf{4} \\ \hline
SPS & BC03 & BC03 & M05 & M05 \\
IMF & Chabrier & Chabrier & Kroupa & Kroupa \\ 
$Z$ & $Z_{\odot}$, $0.4Z_{\odot}$ & $Z_{\odot}$, $0.4Z_{\odot}$ & $Z_{\odot}$, $0.5Z_{\odot}$ & $Z_{\odot}$, $0.5Z_{\odot}$ \\ 
SFH & const+exp & const+del & const+exp & const+del \\
$\tau$ & $0.1, 0.3,$ & $0.1, 0.3,$ & $0.1, 0.3,$ & $0.1, 0.3,$ \\
$\small{\mathrm{[Gyr]}}$ & $0.6$  & $0.6$   & $0.6$    & $0.6$  \\
\hline
\end{tabular}
\end{table}

\section{Number densities and Eddington bias}
\label{appendix_edd}

\begin{figure*}[ht]
    \centering
    \includegraphics[width=1.\textwidth]{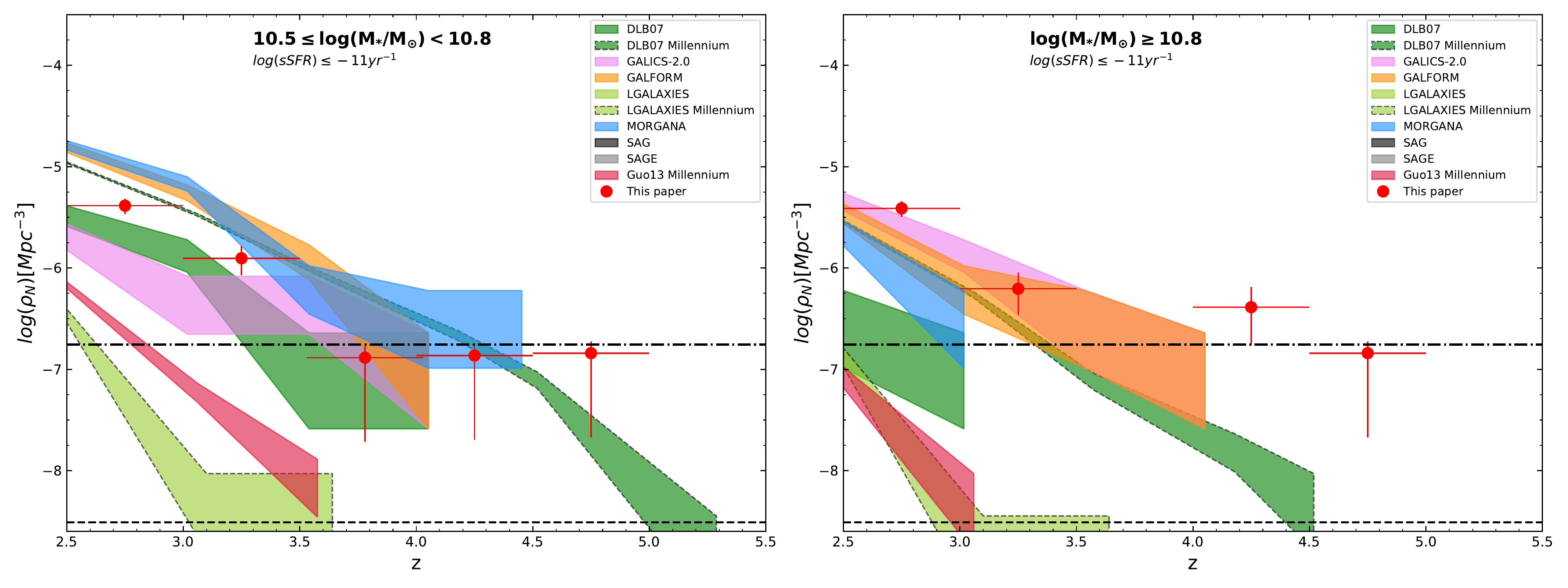}
    \caption{Number densities of QGs in two stellar mass regimes without the convolution in stellar mass. Red dots represent observed galaxies selected in Sec.~\ref{sedfit} from COSMOS with their Poissonian error bars. Colored shaded regions represent $\pm 1\sigma$ data from CARNage and Millennium simulations (same colors as CARNage when the same model is present, but with black dashed lines). The dashed and dash-dotted horizontal black lines represent the number densities expected for 1 object in the simulation volume of Millennium and CARNage respectively.}
    \label{fig:num_dens_noEB}
\end{figure*}

In Fig.~\ref{fig:num_dens_noEB} we present the plot of the number densities of QGs analogous to the one of Fig.~\ref{fig:num_dens}, but without the convolution in stellar mass applied to the last one according to \cite{Ilbert13} to show the effect of the Eddington bias.

\end{appendix}
\end{document}